\documentclass[11pt]{article}
\usepackage{latexsym,graphicx,multirow}
\usepackage{float}
\usepackage{amssymb}
\usepackage{amsmath}
\usepackage{amscd}
\usepackage{amsthm}
\usepackage[left=2cm,top=2.5cm,right=2.5cm,bottom=1.5cm]{geometry}
\usepackage{epstopdf}
\usepackage{cite}
\begin{document}
\begin{center}
	\large{\bf{New Tsallis holographic dark energy with apparent horizon as IR-cutoff in non-flat Universe}} \\
	\vspace{10mm}
	\normalsize{ Nisha Muttathazhathu Ali$^1$, Pankaj$^2$, Umesh Kumar Sharma$^3$, Suresh Kumar P$^4$, Shikha Srivastava$^5$ }\\
	\vspace{5mm}
	\normalsize{$^{1,3,4}$Department of Mathematics, Institute of Applied Sciences and Humanities, GLA University,
		Mathura-281 406, Uttar Pradesh, India}\\
    \normalsize{$^{1,2}$IT Dept.(Math Section), University of Technology and Applied Sciences-HCT,\\
			Muscat-33, Oman}\\
	\normalsize{$^{4}$IT Dept.(Math Section), University of Technology and Applied Sciences-Salalah,\\
			 Oman}\\
	{$^{5}$Department of Mathematics,  Aligarh Muslim University,  Aligarh-202002, Uttar Pradesh, India}\\	 
		 
	\vspace{2mm}
	     $^1$E-mail: nisha.ma123@gmail.com\\
         $^2$E-mail:  pankaj.fellow@yahoo.co.in\\
         $^3$E-mail: sharma.umesh@gla.ac.in\\
		 $^4$E-mail: sureshharinirahav@gmail.com\\
		 $^5$E-mail: shikha.azm06@gmail.com
		\vspace{5mm}
	
		
\end{center}

\begin{abstract}
New Tsallis holographic dark energy  with apparent horizon as IR-cutoff studied by
considering non-flat Friedmann-Lemaitre-Robertson-Walker  metric. The accelerating expansion phase of the Universe is described by using deceleration parameter, equation of state parameter  and density parameter  by using different values of NTHDE parameter ``$\delta$''. The NTHDE Universe's transition from a decelerated to an accelerated expanding phase is described by the smooth graph of deceleration parameter. Depending on distinct values of Tsallis parameter ``$\delta$", we have explored the quintessence behavior of the equation of state parameter. We used Hubble data sets obtained using Cosmic Chronometeric methods and distance modulus measurement of Type Ia Supernova to fit the NTHDE parameters. Stability of our model by analyzing the squared speed of sound investigated as well.
\end{abstract}


\section{Introduction}
The Universe has undergone early-time decelerated and late-time accelerated expansion eras, according to observational data from the last two decades \cite{ref1,ref1a}. There are two major paths to follow in order to explain the reason. Inflation or the dark energy (DE) is the first path, while keeping general relativity as the gravitational theory \cite{ref1b,ref2,ref3,ref3a,ref4}. The second path is to develop expanded and modified gravitational theories based on extra degrees of freedom that trigger acceleration while maintaining general relativity as a particular limit \cite{ref4a,ref4b,ref4c,ref5,ref6,ref8}.\\

The holographic dark energy (HDE) model is one of a series of  DE models that may be used to solve questions regarding the nature of DE \cite{ref8a,ref8b,ref8c,ref8d,ref8e,ref8f,ref8g,ref8h}. In the context of quantum gravity, it is postulated on the grounds of the holographic principle \cite{refN1,refN2}. According to this concept, the number of degrees of freedom in a physical system should scale with its enclosing area rather than it's volume \cite{refN3}. The HDE with apparent and event horizons,  has been widely utilized and their behavior and dependability have been validated using various cosmological parameters. The HDE models based on generalized entropy  can provide substantial explanation for the Universe expansion \cite{ref9,ref9a,ref9b,ref10,ref10a,ref10b,ref10c,ref11,ref12,ref13,ref13a,ref13b,ref13c,ref14,ref14a,ref14b,ref14c,ref14c1}.\\

Tsallis entropy \cite{ref14d,ref14e} is an extension of the Boltzmann-Gibbs entropy to non-extensive systems \cite{ref15,ref16}. It is a generalized entropy measure that has been used to investigate various gravitational and cosmological implications. As a single-parameter  entropy, the generalized entropy is described as \cite{ref15,ref16}
\begin{equation}
\label{eq1}
S_{Q}^{T}=\dfrac{2\exp \left(\dfrac{\delta S_{BH}}{2}\right)\sinh \left(\dfrac{\delta S_{BH}}{2}\right)}{\delta},
\end{equation}
where $\delta$ is an unknown parameter and $S_{BH}$ is the Benkestian entropy. As a result of this entropy and the holographic dark energy theory, a novel model of DE called New Tsallis holographic dark energy (NTHDE) \cite{ref16} is proposed with significant characteristics.\\

Several typical questions arise while studying the Universe's dark energy. Curvature of our Universe is one of the interesting topic. When discussing the evolution of our Universe, the “past, present, and future”, the curvature $k$ plays a crucial part in the Friedmann equation. the three unique possibilities of $k= -1, 0, 1$, which might be “open, flat, or closed”. Due to the involvement of  positive/negative spatial curvature, certain considerations support the hypothesis of non-flat geometry. This problem can be investigated by determining the Universe spatial curvature, which is a quantity that describes the deviation of the Universe's history geometry from flat space geometry \cite{ref17}.
The curvature parameter $\Omega_{k}$, currently quantifies the useful addition mode of spatial curvature to the Universe energy density. The positive and negative values of $\Omega_{k}$ contributes spatially open and closed Universe respectively. The cosmological parameters results  Planck measurements of the CMB anisotropies referred the constraint $\Omega_{k}=-0.044_{-0.015}^{+0.018}$ \cite{ref18}. It is also worth noting that in non-flat geometry, closed Universe models have significantly higher lensing amplitudes than the $\Lambda CDM$ model \cite{ref18}. In either case, the influence of spatial curvature on the evolution of the Universe is the reason why a significant number of researches have been drawn to uncover probable limitations on $\Omega_{k}$ from both current and future cosmic perspectives.\\

From the above studies, the goal of our research is to build a NTHDE in a non-flat Universe with an IR-cutoff as the apparent horizon. After introduction the basic equations of NTHDE in a non-flat Friedmann-Lemaitre-Robertson-Walker (FLRW) Universe are derived. Through section 3 cosmic evolution in view of equation of state and density parameter (DP) are described. Section 4 is dedicated to the data analysis by using observational Hubble and SNIa data sets. The stability of the model is discussed under section 5. At the end, the result concluded in section 6.

\section {NTHDE in non-flat Universe} 
The Universe is isotropic and homogeneous in large scale. The FLRW metric is the greatest match to describe the non-flat Universe's geometry. It's equation is:
\begin{equation}
\label{eq2}
ds^{2}= -dt^{2}+a^{2}(t)\left[\dfrac{dr^{2}}{1-kr^{2}}+r^{2}d\Omega^{2}\right],
\end{equation}
where $ a(t)$  is the scale factor of the Universe and spatial curvature $k=-1,0,1$ corresponds to closed, flat and open Universe respectively.\\

Since the HDE hypothesis together with the Tsallis entropy content \cite{ref19} of black hole claims that if the present accelerated Universe handled by the energy of vacuum ($\rho_{\Lambda}^{T}$), then the amount of stored energy in a $L^3$ size box must be less than the same size black hole energy \cite{ref16}, leads to get the vacuum energy $\rho_{\Lambda}^{T}$ as
\begin{equation}
\label{eq3}
\Lambda^4\equiv\rho_{\Lambda}^{T}\propto\frac{S_{Q}^{T}}{L^4},
\end{equation}
In our study we considered apparent horizon as the IR-cutoff, then the horizon length $L$ or radius $\tilde{r}_{A}$ is given by \cite{ref20,ref21,ref22,ref23},
\begin{equation}
\label{eq4}
\tilde{r}_{A}=\dfrac{1}{\sqrt{\frac{k}{a^{2}}+H^{2}}},
\end{equation}
hence we get the NTHDE density as 
\begin{equation}
\label{eq5}
\rho_{\Lambda}^{T}=\dfrac{2D^{2}\rho_{c}}{X}\left(\Omega_{k}+1\right)^{2}\exp \left(\dfrac{X}{2\left(\Omega_{k}+1\right)}\right)\sinh \left(\dfrac{X}{2\left(\Omega_{k}+1\right)}\right),
\end{equation}
 where $D^2=\pi T^2$ ($T^2$ is constant), $\rho_{c}=\dfrac{3H^2}{8\pi}$ and $X=\dfrac{\delta\pi}{H^2}$ with $\delta$, known in the most recent research as Tsallis parameter. Obviously equation (\ref{eq5}) correlates with NHDE in a flat Universe for the case $\Omega_{k}=0$ \cite{ref16}. For the described metric, by including NTHDE density and dark matter density Friedmann's equations adopt the following forms
\begin{equation}
 \label{eq6}
 H^{2}+\dfrac{k}{a^{2}}=\dfrac{8\pi}{3}G\left(\rho_{\Lambda}^{T}+\rho_{m}\right),
\end{equation}
\begin{equation}
 \label{eq7}
 H^{2}+\dfrac{2\dot{H}}{3}+\dfrac{k}{3a^{2}}=\dfrac{-8\pi}{3}Gp_{\Lambda}^{T},
\end{equation}
 where $p_{\Lambda}^{T}$ is the pressure of NTHDE. We describe the non-relativistic matter-energy DP ($\Omega_{m}$), dark energy DP ($\Omega_{D}$) and spatial curvature DP ($\Omega_{k}$) as 
 \begin{equation}
 \label{eq8}
 \Omega_{m}=\dfrac{8\pi G}{3H^{2}}\rho_{m},\qquad \Omega_{D}=\dfrac{8\pi G}{3H^{2}}\rho_{\Lambda}^{T},\qquad \Omega_{k}=\dfrac{k}{a^{2}H^{2}}.
\end{equation}
As a result, the equation  can be rewritten as
\begin{equation}
 \label{eq9}
 \Omega_{k}+1=\Omega_{m}+\Omega_{D}.
\end{equation}
The conservation law for NTHDE and matter are defined as
\begin{equation}
 \label{eq10}
 \dot{\rho}_{\Lambda}^{T}+3\left(1+\omega_{\Lambda}^{T}\right)\rho_{\Lambda}^{T}H=0,
\end{equation}
\begin{equation}
 \label{eq11}
 \dot{\rho}_{m}+3\rho_{m}H=0,
\end{equation}
where $\omega_{\Lambda}^{T}=\dfrac{p_{\Lambda}^{T}}{\rho_{\Lambda}^{T}}$ is the equation of state(EoS) of NTHDE. To derive the behavior of NTHDE, differentiate the equation (\ref{eq5}) with respect to the cosmic time $t$, we get
\begin{equation}
 \label{eq12}
 \dot{\rho}_{\Lambda}^{T}=2H^{3}\left(\frac{\dot{H}}{H^{2}}-\Omega_{k}\right)\left(\frac{2\rho_{\Lambda}^{T}}{H^{2}(\Omega_{k}+1)}\left(1-\frac{X}{4(\Omega_{k}+1)}\right)-\frac{3T^{2}}{8}\exp\left(\frac{X}{2(\Omega_{k}+1)}\right)\cosh\left(\frac{X}{2(\Omega_{k}+1)}\right)\right).
\end{equation}
Now the equation (\ref{eq7}) and equation (\ref{eq8}) together provides
\begin{equation}
 \label{eq13}
 \dfrac{\dot{H}}{H^{2}}=\dfrac{-3}{2}\left(\omega_{\Lambda}^{T}\Omega_{D}+\frac{\Omega_{k}}{3}+1\right),
\end{equation}
hence the equation for deceleration parameter($q$) is discovered as
\begin{equation}
 \label{eq14}
 q=-\dfrac{\dot{H}}{H^{2}}-1=\dfrac{1}{2}\left(3\omega_{\Lambda}^{T}\Omega_{D}+\Omega_{k}+1\right).
\end{equation}
Making use of equations (\ref{eq5}), (\ref{eq10}) and (\ref{eq12}), we get the EoS parameter($\omega_{\Lambda}^{T}$) as
 \begin{equation}
 \label{eq15}
 \omega_{\Lambda}^{T}=\dfrac{1}{3}\left(\dfrac{\delta\pi(\dot{H}+H^{2})\left(1+\coth\left(\frac{X}{2(\Omega_{k}+1)}\right)\right)}{H^{4}(\Omega_{k}+1)^{2}}-\dfrac{\delta\pi\left(1+\coth\left(\frac{X}{2(\Omega_{k}+1)}\right)\right)+4(\dot{H}+H^{2})}{H^{2}(\Omega_{k}+1)}+1\right).
\end{equation}
In light of equations (\ref{eq5}) and (\ref{eq8}), the DP of NTHDE can also be written as
\begin{equation}
 \label{eq16}
 \Omega_{D}=\dfrac{2T^{2}H^{2}}{\delta}\left(\Omega_{k}+1\right)^{2}\exp \left(\dfrac{X}{2\left(\Omega_{k}+1\right)}\right)\sinh \left(\dfrac{X}{2\left(\Omega_{k}+1\right)}\right).
\end{equation}
\section{Evolution of cosmological parameters}
In this part, we will look at the cosmological behavior in a open and closed Universes where the DE sector is the NTHDE. The behavior of cosmological evolution is obtained in terms of the red shift $z$ by setting initial scale value as $a_{0}=1$. We numerically develop equations, assuming the initial parameters $\Omega_{D}=0.70$ and $H_{0}=67.9$, which are consistent with recent results \cite{ref18}. Also  for an open Universe ($\Omega_{k}(0)=0.0026$) and closed Universe ($\Omega_{k}(0)=-0.0012$) we used different observational best-fit values of NTHDE model parameter as $\delta=1400, 1450$, $1500$ and $T=0.35$.\\

\begin{figure}[H]
		\centering
		\includegraphics[scale=0.6, angle=0]{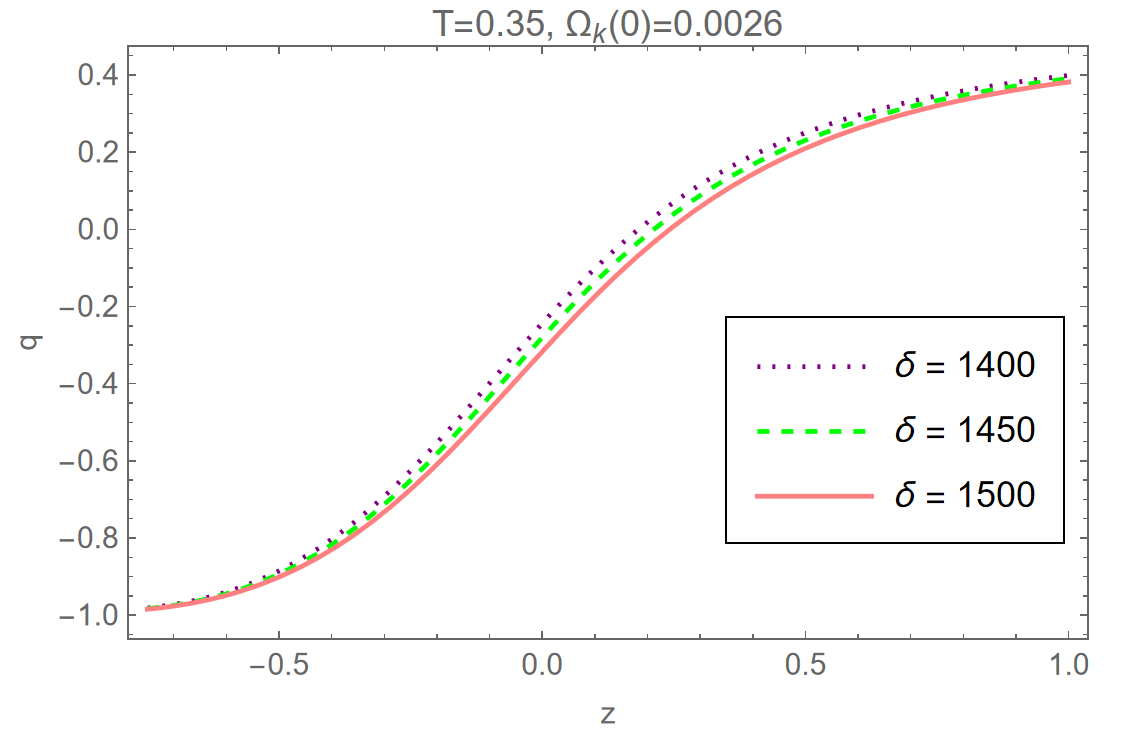}
		\caption{The deceleration parameter behavior for open Universe with $\Omega_{k}(0)=0.0026$ , $T=0.35$, $H_{0}=67.9$ and different $\delta$ values.}
		\label{fig1}
		\end{figure}

\begin{figure}[H]
		\centering
		\includegraphics[scale=0.6, angle=0]{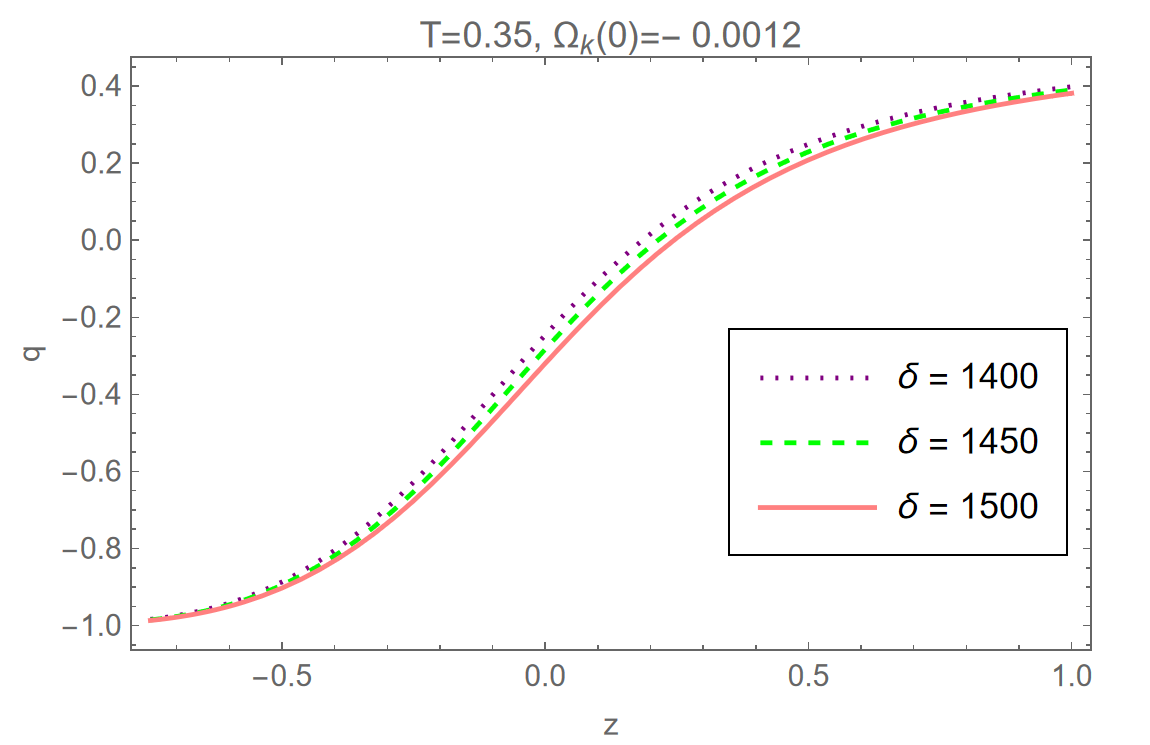}
		\caption{The deceleration parameter behavior for closed Universe with $\Omega_{k}(0)=-0.0012$ , $T=0.35$, $H_{0}=67.9$ and different $\delta$ values.}
		\label{fig1a}
		\end{figure}

From figure \ref{fig1} and figure \ref{fig1a}, we illustrate the numerical solution of $q$ with the initial conditions which are mentioned in the above paragraph. We noted that $q$ shows a Universe expanding at a faster rate, and also can observe that the NTHDE model is approaching the accelerating phase around the red shift $-0.6<z<0.2$. For the open and closed Universe, we can see that $q\rightarrow-1$, when $z\rightarrow-1$. The cosmos is transitioning from deceleration to acceleration at the late stage in its evolution.\\
		
\begin{figure}[H]
		\centering
		\includegraphics[scale=0.6, angle=0]{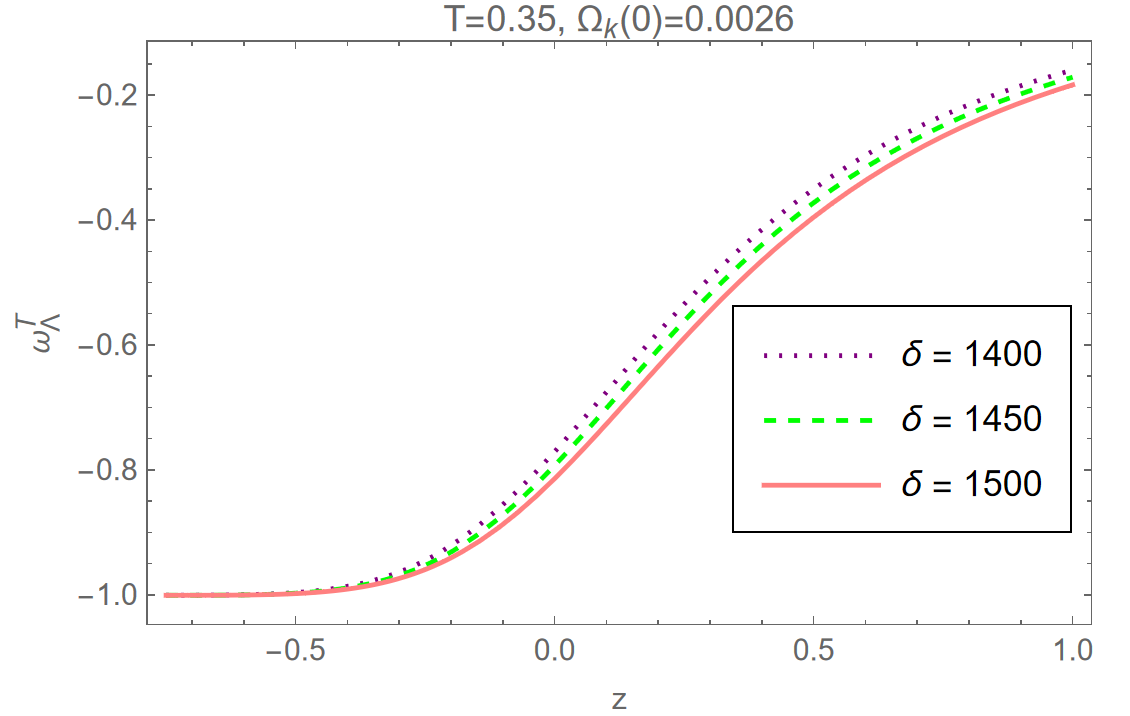}
		\caption{The EoS parameter($\omega_{\Lambda}^{T}$) behavior for open Universe $\Omega_{k}(0)=0.0026$ with $\Omega_{D}(0)=0.70$, $T=0.35$, $H_{0}=67.9$ and different $\delta$ values.}
		\label{fig2}
		\end{figure}

\begin{figure}[H]
		\centering
		\includegraphics[scale=0.6, angle=0]{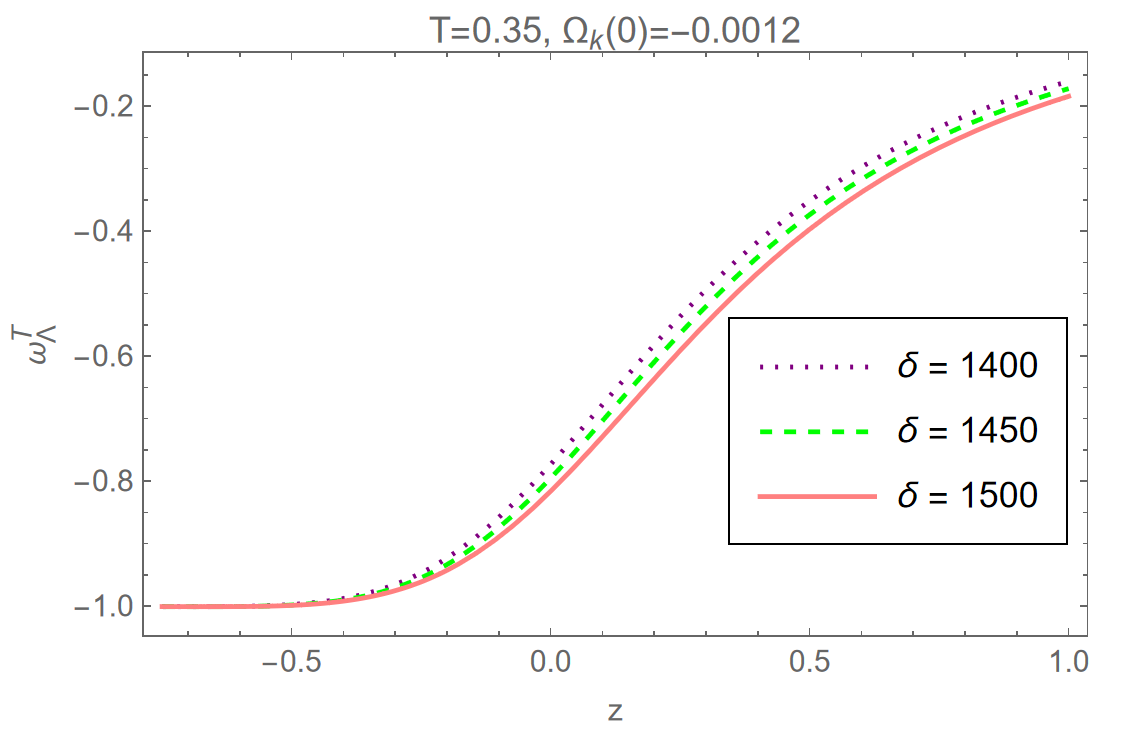}
		\caption{The EoS parameter($\omega_{\Lambda}^{T}$) for closed Universe $\Omega_{k}(0)=-0.0012$ with $\Omega_{D}(0)=0.70$, $T=0.35$, $H_{0}=67.9$ and different $\delta$ values.}
		\label{fig2a}
		\end{figure}

The major task in observational cosmology is estimating the EoS parameter for DE. In figure \ref{fig2} and figure \ref{fig2a}, the EoS parameter's behavior versus ``$z$" is presented for open and closed Universe respectively. We observed that the EoS parameter is approaching to $-1$ at low redshifts. As a result, the NTHDE acts similarly like vaccum dark energy . Our model is good agreement with the $\Lambda CDM$ as $\omega_{\Lambda}^{T}\rightarrow -1$ at $z\rightarrow-1$. For open and closed Universes we can observe that the EoS parameter does not cross $\omega_{\Lambda}^{T}=-1$ at different values of $\delta$. Also $\omega_{\Lambda}^{T}$ completely lies in the quintessence region ($\omega_{\Lambda}^{T}>-1$). We will be able to calculate the asymptotic value of $\omega_{\Lambda}^{T}$ at late time. This suggest that in $\Lambda CDM$ cosmology, our $\omega_{\Lambda}^{T}$ has an interesting behavior.
		
\begin{figure}[H]
		\centering
		\includegraphics[scale=0.6, angle=0]{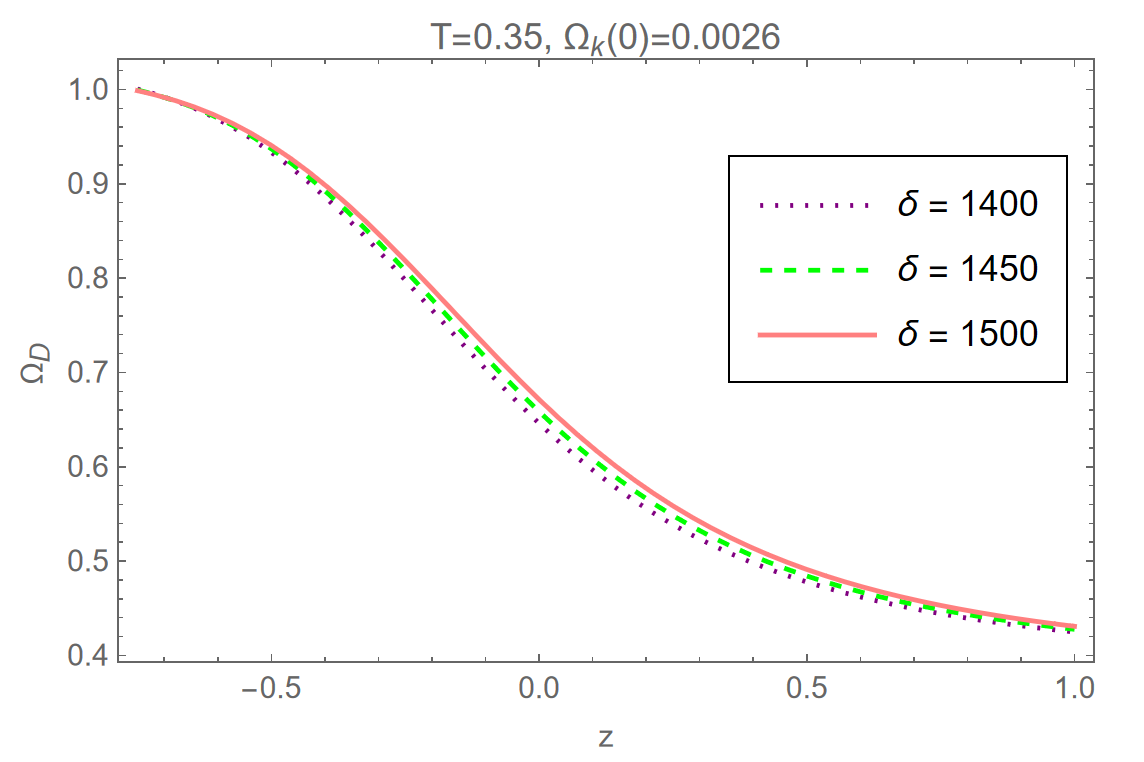}
		\caption{The density parameter($\Omega_{D}$) behavior for open Universe $\Omega_{k}(0)=0.0026$ with $\Omega_{D}(0)=0.70$, $T=0.35$, $H_{0}=67.9$ and different $\delta$ values.}
		\label{fig3}
		\end{figure}

\begin{figure}[H]
		\centering
		\includegraphics[scale=0.6, angle=0]{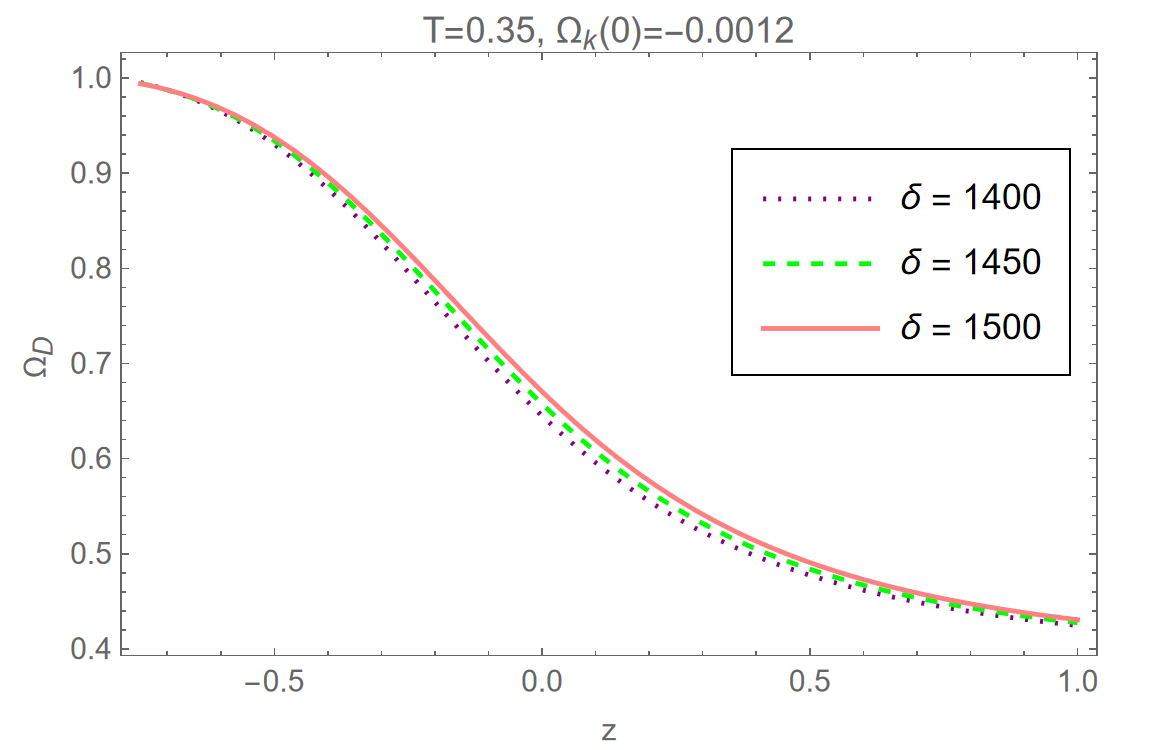}
		\caption{The density parameter($\Omega_{D}$) behavior for closed Universe $\Omega_{k}(0)=-0.0012$ with $\Omega_{D}(0)=0.70$, $T=0.35$, $H_{0}=67.9$ and different $\delta$ values.}
		\label{fig3a}
		\end{figure}

The evolution of DP ($\Omega_{D}$) is shown in figure \ref{fig3} and figure \ref{fig3a} for open and closed Universe at different values of NTHDE parameter $\delta$ respectively. We have taken $\Omega_{D}(0)=0.70$ as initial condition. In the figures it is very clear that $\Omega_{D}\rightarrow1$ at $z\rightarrow-1$ implies the dark energy domination in the future. It is interesting that, for greater transparency, cosmic evolution should be extended to late periods,  so that  we can witness the results of cosmic evolution in total DE dominance, as predicted.
		
		\section{ Observational Hubble Data Analysis }
\begin{figure}[H]
		\centering
		\includegraphics[scale=0.55, angle=0]{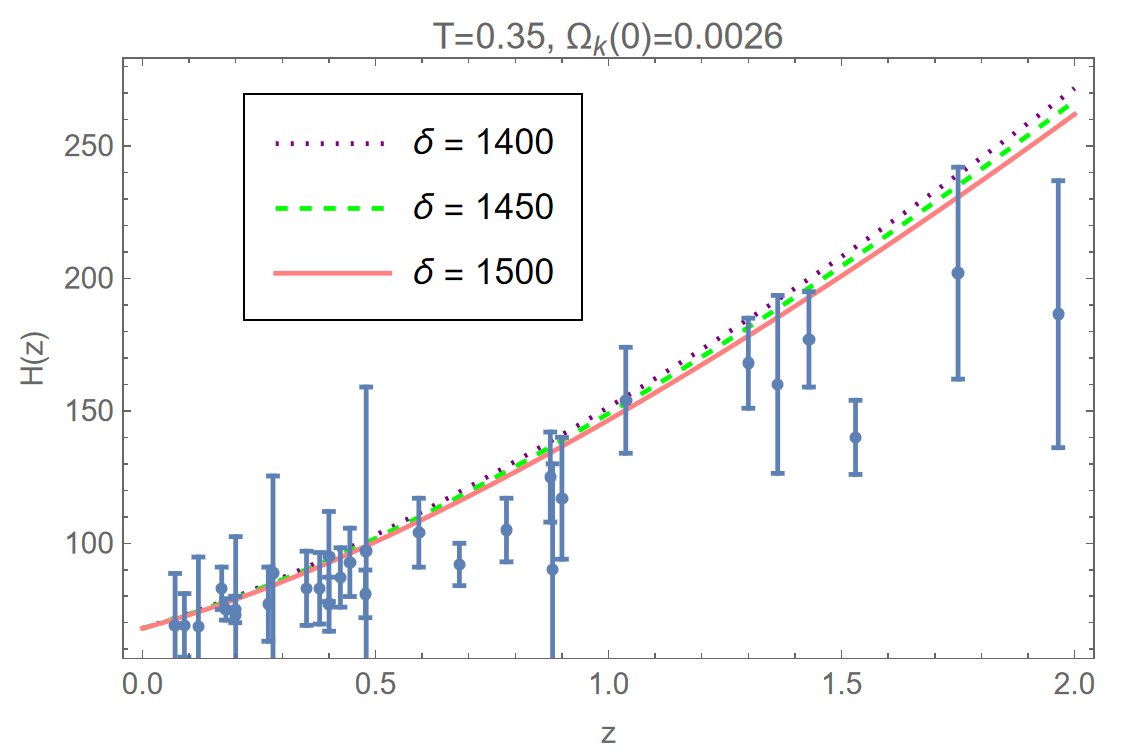}
		\caption{The Hubble parameter($H(z)$) behavior for open Universe $\Omega_{k}(0)=0.0026$ with $\Omega_{D}(0)=0.70$, $T=0.35$, $H_{0}=67.9$ and different $\delta$ values.}
		\label{fig4}
		\end{figure}

\begin{figure}[H]
		\centering
		\includegraphics[scale=0.55, angle=0]{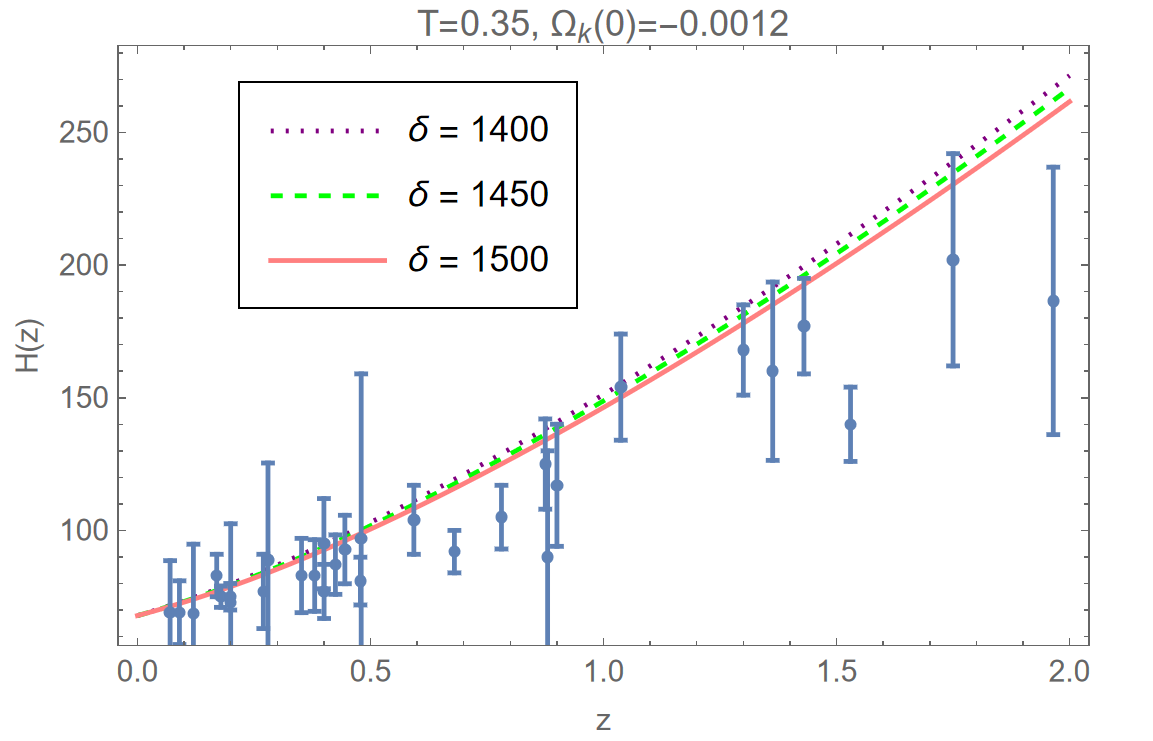}
		\caption{The Hubble parameter($H(z)$) behavior for closed Universe $\Omega_{k}(0)=-0.0012$ with $\Omega_{D}(0)=0.70$, $T=0.35$, $H_{0}=67.9$ and different $\delta$ values.}
		\label{fig4a}
		\end{figure}

Since this observational Hubble data is based on model-independent direct findings, it may be used to limit cosmological parameters. This has shown to be helpful in comprehending the nature of Dark energy. In our NTHDE model, we use the current data obtained via Cosmic Chronometer mode. This mode of data depend on the method of distinct dating of galaxies. For our data analysis we considered the $30 H(z)$ observational data set from table 4 of \cite{ref24} which is in the redshift interval of $0.07\leqslant z \leqslant1.965$. For both open and closed Universes  by using figure \ref{fig4} and figure \ref{fig4a}, we described the Hubble parameter($H(z)$) evolution for our NTHDE model and it's comparison with Hubble data sets. We can conclude that NTHDE in a non-flat Universe agrees perfectly with Hubble data.
		
      \subsection{ Observational SNIa Data Analysis }
\begin{figure}[H]
		\centering
		\includegraphics[scale=0.6, angle=0]{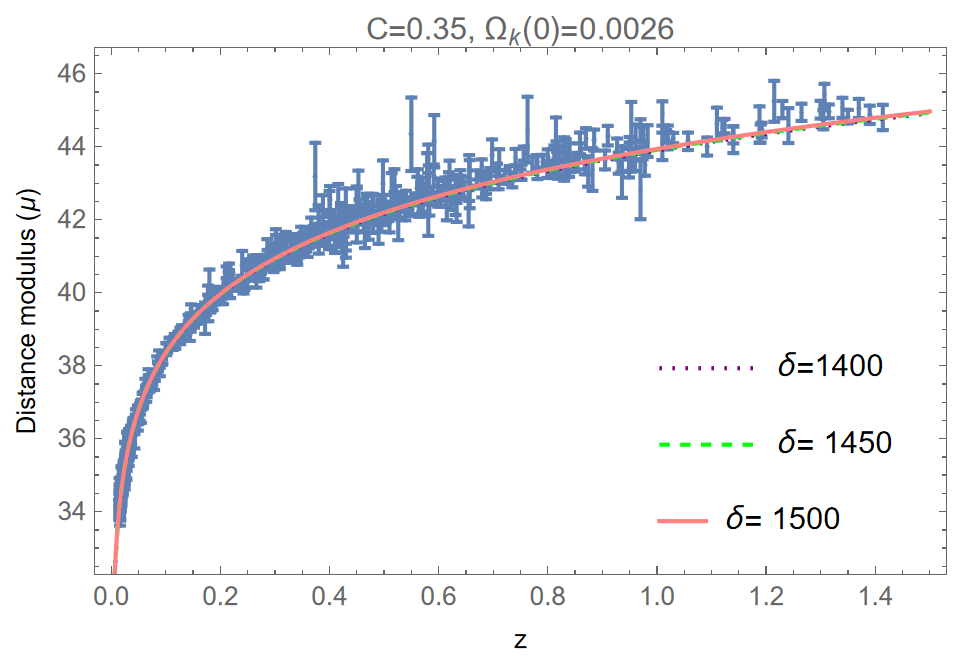}
		\caption{The distance modulus($\mu$) behavior for open Universe $\Omega_{k}(0)=0.0026$ with $\Omega_{D}(0)=0.70$, $T=0.35$, $H_{0}=67.9$ and different $\delta$ values.}
		\label{fig5}
		\end{figure}

\begin{figure}[H]
		\centering
		\includegraphics[scale=0.6, angle=0]{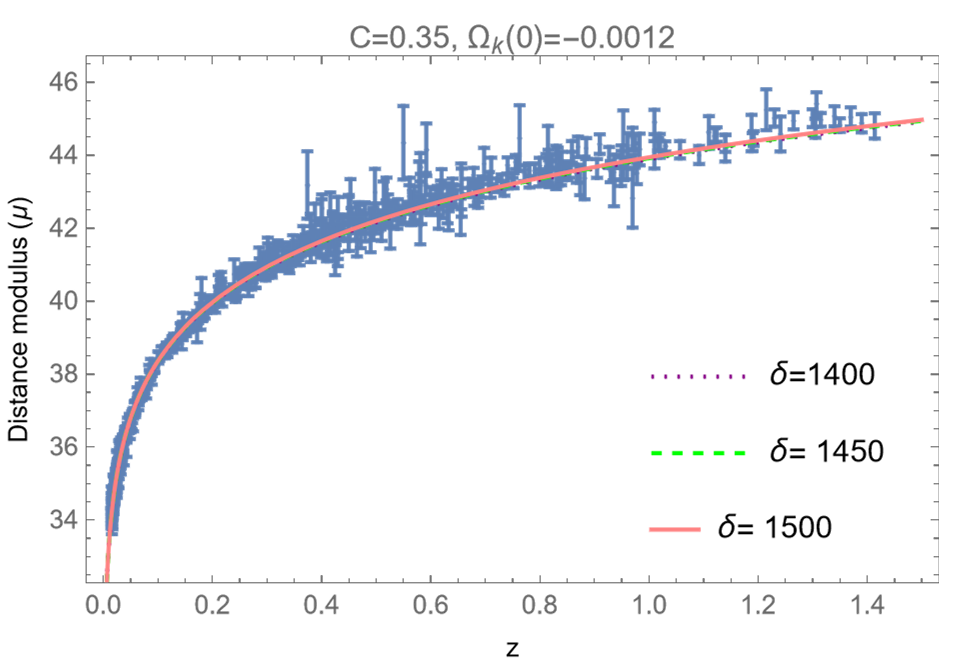}
		\caption{The distance modulus($\mu$) behavior for closed Universe $\Omega_{k}(0)=-0.0012$ with $\Omega_{D}(0)=0.70$, $T=0.35$, $H_{0}=67.9$ and different $\delta$ values.}
		\label{fig5a}
		\end{figure}

 The studies of nearby and distant Type Ia Supernovae (SNIa) revealed that the Universe's expansion is accelerating up during this time \cite{ref1a,ref26,ref27,ref28}. The apparent luminosity of SNIa are compared over a range of redshifts to estimate cosmological parameters. As a result, we utilized the distance modulus data set sample of 580 Union 2.1 scores in association with SNIa \cite{ref29}. The redshift and luminosity distance relationship is the most well-known approach to observe the expansion of the Universe \cite{ref30,ref3}. The redshift of light emitted by a faraway object due to the expansion of the cosmos is a well-measured quantity. In terms of redshift($z$), we may find the equation for the luminosity distance($D_{L}$). The $D_{L}$ to the object with the help of the flux of the source in terms of $z$ and radial coordinate($r$) is given by
 \begin{equation}
 \label{eq17}
D_{L}=a_{0}r(z+1),
\end{equation}
here $r$ stand for radial coordinate of the source. From \cite{ref3}, we can get the formula for $D_{L}$ as
 \begin{equation}
 \label{eq18}
D_{L}=\dfrac{c\left(1+z\right)}{H_{0}}\int_{0}^{z}\frac{dz}{h(z)}, \qquad h(z)=\frac{H}{H_{0}},
\end{equation}
 SNIa are so luminous that they can be seen at extremely high redshifts. They have a similar brightness that is independent of redshift and accurately calibrated by their light curves. As a result, they make excellent standard candles for measuring brightness distances. Distance modulus ($\mu$) are used to represent luminous distance data, is given by \cite{ref3}
  \begin{equation}
 \label{eq19}
\mu=25+5\log \left(\dfrac{D_{L}}{M_{pc}}\right).
\end{equation}
By combining equations (\ref{eq18}) and (\ref{eq19}), the expression for $\mu$ is given by
 \begin{equation}
 \label{eq20}
\mu=25+5\log \left[\dfrac{c\left(1+z\right)}{H_{0}}\int_{0}^{z}\frac{dz}{h(z)}\right].
\end{equation}
Figure \ref{fig5} and figure \ref{fig5a} describe the evolution of $\mu$ versus $z$ and it is compared with SNIa data \cite{ref29}. In the figures it is obvious that, there is a  strong reflection of the NTHDE model on the observed values of $\mu$ for each data set, which supports our model.
		
\section{Stability}
\begin{figure}[H]
		\centering
		\includegraphics[scale=0.6, angle=0]{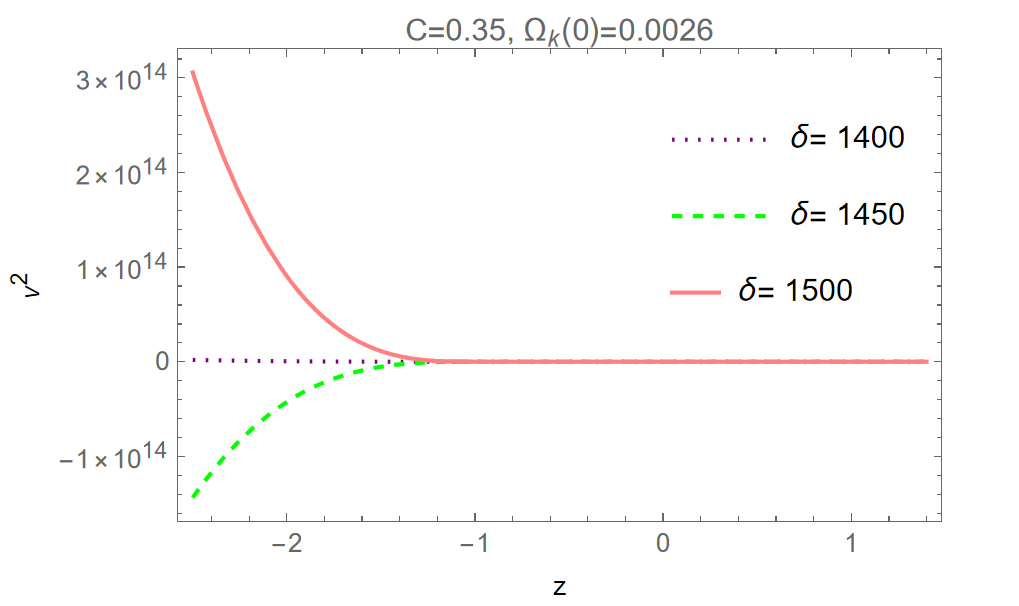}
		\caption{The squared speed of sound ($v^{2}$) behavior for open Universe $\Omega_{k}(0)=0.0026$ with $\Omega_{D}(0)=0.70$, $T=0.35$, $H_{0}=67.9$ and different $\delta$ values.}
		\label{fig6}
		\end{figure}

\begin{figure}[H]
		\centering
		\includegraphics[scale=0.6, angle=0]{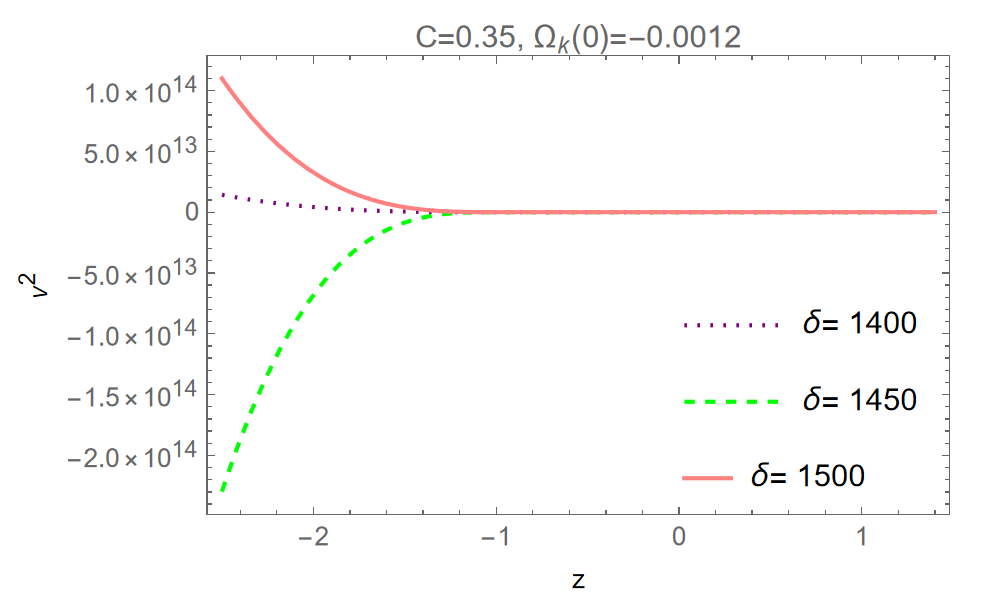}
		\caption{The squared speed of sound ($v^{2}$) behavior for closed Universe $\Omega_{k}(0)=-0.0012$ with $\Omega_{D}(0)=0.70$, $T=0.35$, $H_{0}=67.9$ and different $\delta$ values.}
		\label{fig6a}
		\end{figure}

The current section looks at the classical stability of the NTHDE model using the squared speed of sound for non-flat FLRW Universe. The squared speed of sound ($v^{2}$) is defined as \cite{ref32, ref33} 
\begin{equation}
 \label{eq21}
v^2=\dfrac{dp_{D}}{d\rho_{\Lambda}^{T}}=\dfrac{\rho_{\Lambda}^{T}}{\dot{\rho_{\Lambda}^{T}}}\dot{\omega_{\Lambda}^{T}}+\omega_{\Lambda}^{T}.
\end{equation}
Using several values of parameter $\delta$, we plotted the behavior of $v^{2}$ versus $z$ in figure \ref{fig6} and figure \ref{fig6a}. The instability of the trajectory against background perturbation is shown by its negative value. Therefore from the figure we can conclude that at $\delta=1450$ (green dotted curve), we have $v^{2}<0$ and our NTHDE model is not stable in non-flat Universe. But at $\delta=1400$ and $\delta=1500$ (red solid curve and purple dotted curve respectively) $v^{2}>0$, our NTHDE model is stable in both open and closed Universes at late times. As a result, the new Tsallis parameter's permitted values are constrained by the model's classical stability requirement.
		
		\section{Conclusion}
In this study, we looked at the NTHDE model with the apparent horizon as the IR-cutoff. $T$ and $\delta$ are the two main parameters that determine this model. The Tsallis entropy, which was involved in the normal Benkestian entropy, was utilized to discuss NTHDE in this work. We used the observational data set of $H(z)$ collected by Cosmic Chronometric method and data set taken from SNIa to restrict the parameters.\\

The NTHDE model in open and closed Universes leads to an interesting cosmic phenomenology. The transition from the early deceleration phase to the current accelerating phase of the NTHDE appears to be smooth, according to the graph of deceleration parameter ($q$). For distinct $\delta$ values, $\omega_{\Lambda}^{T}$ of NTHDE lies in the quintessence region. It was discovered to be lies in the $\omega_{\Lambda}^{T}\geq -1$ region, which is a good line with the accelerating Universe. In our model $\omega_{\Lambda}^{T}$ does not cross $-1$. The dark energy dominated cosmos can be proved at late times based on the density parameter graph. We discovered that the evolution of Hubble parameter and distance modulus are in good accord with the cosmic chronometric and SNIa observational data sets. Using the squared of sound speed ($v^{2}$), we demonstrated that the NTHDE model satisfies the criteria of stability at late times in both open and closed Universes.\\

Moreover, in addition to constrain the new parameters more precisely, it is necessary to extend the current study using data from Baryon Acoustic Oscillation, Pantheon SNIa, and Cosmic Microwave Background observations. This field of research is currently being pursued and more findings will be revealed in future studies.

\end{document}